\documentclass[preprint,12pt]{elsarticle}

\usepackage{graphicx}
\usepackage{amsmath,amssymb,latexsym}
\usepackage[colorlinks, plainpages]{hyperref}
\usepackage[english,francais]{babel}

\newtheorem{theorem}{Theorem}[section]

\newtheorem{definition}[theorem]{Definition}

\newcommand{\be}{\begin{equation}}
\newcommand{\ee}{\end{equation}}
\newcommand{\beq}{\begin{eqnarray}}
\newcommand{\eeq}{\end{eqnarray}}

\begin{document}

\begin{frontmatter}

\title{Hearing  the   Symmetries of Crystal Lattices from the Integrated Acoustic Spectrum}

\author{H. Mohades and B. Honari \fnref{}}
%\ead{hoseinmohades@aut.ac.ir}
%\author{B. Honari\corref{cor1}\fnref{label2}}
%\ead{mmreza@aut.ac.ir}
%\address[label1,label2]{Department of Pure Mathematics, Faculty of Mathematics and Computer Science, Amirkabir University of Technology, No. 424, Hafez Ave.\fnref{label3}}

\begin{abstract}
Let $C$ be a crystal and $\phi$ be a periodic realization of it in $\mathbb{R}^n$, also let $L$ be the lattice group of   $\phi(C)$ which preserves the covering space nature of crystal lattices. In this article, firstly, we define the concept of  acoustic spectrum of the crystal lattice C and then provide the  algebraic formalism of the question of  finding the frequencies  of the torus $\mathbb{R}^n/L$, when the set of   acoustic spectrum is known. An answer for crystals with uniform atomic force constants is given.

\end{abstract}


\end{frontmatter}



\section{Introduction}
Laplace-Beltrami operator is a natural second order elliptic operator on a Riemannian manifold defined as div $\circ$ grad. It is well known that on a closed manifold, this operator has discrete positive eigenvalues with finite multiplicities [4].Two Riemannian manifolds are isospectral if  their Laplace-Beltrami operator have the same spectrum, considering multiplicities. A fundamental question by Mark Kac asks whether it is possible to find two nonisometric isospectral manifolds. The first  answer to this question was provided by Milnor's 16 dimensional tori that is a geometric realization of self-dual lattices with the same theta functions \cite{mi}. Naturally, lattices appear in the theory of crystallography as symmetries of a crystal lattice. We mean by a crystal lattice, a periodic harmonic realization of a commutative covering space of a finite graph. In the nature, the interatomic forces  lead to  oscillations of  crystal's atoms around their equilibrium positions.  These oscillations are called crystal lattice vibrations. Physicists  usually  decompose the system of oscillations into independent simple harmonic oscillators,
and calculate the distribution of vibration frequencies \cite{ ss}. This method is the same as the theory of Fourier series for a vibrated chord. Acoustic phase velocities are the phase velocity of elastic waves in the uniform elastic body
corresponding to the crystal lattice.
In this article we consider the integration of acoustic phase velocity in the direction of closed geodesics of the lattice character group and we find the algebraic formalism of hearing the eigenvalues of Laplacian (or equivalently the frequencies) on the torus $\mathbb{R}^n/L$.

This paper is organized as follows.
In section 2 a review of the notion of a crystal lattice and its realization is presented. Section 3 is devoted to the theory of vibrations of a crystal lattice. Finally, in section 4 the algebraic formalism of hearing the eigenvalues of Laplacian  on the torus  $\mathbb{R}^n/L$  is provided and some special cases are studied.
\section{Crystal Lattices}
In this section we follow the Sunada's graph theory method to introduce the notion of a crystal graph \cite{su}.
\subsection{Graphs and crystals} A graph is an ordered pair $X = (V,E)$ of disjoint sets $V$ and $E$ with two
maps $o:E\rightarrow V$ and $t:E\rightarrow V$. It is finite if both $V$ and $E$ are finite sets.  A geometric graph is $V\cup (E\times[0,1])/\sim$ where the equivalence relation $\sim$ is defined by $o(e) \sim (e,0), t(e) \sim (e,1)$. Let $X$ and $X_0$ be two geometric graphs and let $\pi:X\rightarrow X_0$ be a covering map. The graph $X$ is called an abelian covering space of  $X_0$ if  the deck transformation group is abelian.
An abstract  crystal $C$ is an infinite regular covering of a geometric graph $X$ over a finite graph $X_0$, with
free abelian deck transformation group. Every abstract crystal is obtained by choosing a subgroup $H$ of the homology group $H_1(X_0,\mathbb{Z})$ when $\frac{H_1(X_0,\mathbb{Z})}{H}$ is a free abelian group.
\subsection{Realization} Set $l^2(V)=\{f:V\rightarrow \mathbb{C}|\sum_{x\in V}|f(x)|^2<\infty\}.$ \\ \begin{definition}The discrete Laplacian $\Delta: l^2(V)\rightarrow l^2(V)$ is defined by%\frac{1}{m_x}
	\begin{equation}\Delta(f)(x)=\sum_{e\in E,o(e)=x}(f(t(e))-f(o(e))).\end{equation} \end{definition}
\begin{definition}(Periodic realization) A piecewise linear map $\phi:X\rightarrow \mathbb{R}^n$ is said to be a periodic realization,
	if there exists an injective homomorphism $\rho:L\rightarrow \mathbb{R}^n$  such  that\\ a) $\phi(\sigma x)=\phi(x)+\rho(\sigma)  (x \in V, \sigma \in L)$ and b) $\rho(L)$ is a lattice subgroup of $\mathbb{R}^n$.\end{definition}

The periodic realization $\phi:X\rightarrow\mathbb{R}^n $ is harmonic (or standard in the Sunada's notation) provided that it is a solution
of the discrete Laplace equation $\Delta \phi= 0$ and there exists a positive constant $c$ such that \begin{equation}\label{se}
	\sum_{e\in X_0}x.(\phi(t(e'))-\phi(o(e'))(\phi(t(e'))-\phi(o(e'))=cx \, \,, \forall x\in \mathbb{R}^n
\end{equation}where $e' \in \pi^{-1}(e)$ is arbitrary (maximal orthogonality property).

\section{Vibration of lattices}
Harmonic realization of a lattice is the state of  minimum energy of its realizations which depends on the function $\rho:L=\frac{H_1(X_0,\mathbb{Z})}{H}\rightarrow R^n$ as the symmetries of the covering space $\pi:X\rightarrow X_0$.
In crystallography it is assumed that two elements (atoms)  in the same  orbit of $\rho(L)$ are of the same type. At temperatures close to zero a crystal lattice vibrates about its equilibrium position (its harmonic realization) by effect of its inter-atomic forces. The  motion $f$ satisfies the equation \begin {equation} \frac{d^2(f)}{dt^2}=Df\end{equation}
where $D$ is the discrete elastic Laplacian defined by \begin{equation} Df (x) =\frac{1}{m(x)}\sum_{e\in E_x}A(e)(f(t(e))-f(o(e)))\end{equation}
for positive definite symmetric $m\times m$ matrices $A(e)$(this is an extra, but useful condition), $ m=card(E_0) $,  and masses $m(x)$ associated to atoms $x\in V$. Physicists call $A$,  the matrix of atomic force constants.
$D$
is an $L$-equivariant linear bounded self-adjoint
operator on the  Hilbert space generated by the space $C(V,\mathbb{C}^n)$  equipped with the inner product $<f,g>=\sum_{x\in V} f(x).\overline{g(x)}m(x)$. %The operator $ D$ is called the discrete elastic Laplacian.

\subsection {Hamiltonian formalism of the motion equation} In this section a Hamiltonian formalism is used to provide a  decomposition of a vibration to simpler harmonic vibrations \cite{ss}.
Let $w:l^2(V,m)\times  l^2(V,m)\rightarrow \mathbb{R}$ be a symplectic form on
$l^2(V,m)=\{\sum_{x\in V}|f(x)|^2m(x)<\infty\}$ defined by $w(u,v)=Im<u,v>=Im( \sum_{x\in V}u(x)v(x)m(x))$. Let $H(u) =\frac{1}{2}<(-D)^{\frac{1}{2}}u, u>$.
Let $ \widehat{L}$ be the unitary character group of $ L$, and $d\chi$ denotes the normalized Haar
measure on $ \widehat{L}$. Then
the Bloch decomposition of the Hamiltonian system $(l^2(V,m),w,H)$ is
\begin{equation}
	(l^2(V,m),w,H)=\int_{\widehat{L}}^\bigoplus(l^2(V,m)_\chi,w_\chi,H_\chi)d\chi \,,\end{equation} where $l^2(V,m)_\chi=\{u\in l^2(V,m)|u(\sigma x)=\chi(\sigma)u(x)\,, \forall \sigma \in L\}$. In fact, the elements $\chi \in \widehat{L} $ are the same as the angular frequencies. Using the index $\chi$ for an operator means that we restrict its domain to the  vector space $l^2(V,m)_\chi$.
Note that we have a decomposition of $D$ as $\int_{\widehat{L}}^\bigoplus D_\chi  d\chi$. The operators $-D_\chi , \, \chi\ne 1$ are positive definite.
Zero is an eigenvalue of $−D_1$ of multiplicity $n$ with constant eigenfunctions.\\
For the eigenmodes (eigenfunctions) corresponding to the first $n$ eigenvalues of $D_\chi$, neighboring atoms move in phase with each other with the same amplitude and each of which is called an $\mathbf{acoustic\, phase}$. The velocity of these phases is called the acoustic phase velocity.
It is equal to $\frac{1}{2\pi\parallel{\chi}\parallel}s_i(\chi),\, i=1,...,n$, where one can prove
\begin{theorem}\label{3} $s_i(\chi)^2\,\,\,(i = 1, \dots, n)$ are eigenvalues of the symmetric
	matrix
	\begin{equation}A_\chi :=
		\frac{2\pi^2}{
			m(V_0)}
		\sum_{e\in E_0}(\chi.v(e))^2A(e).\end{equation}
	
	In particular, $s_i(\chi)^2 > 0$ for $\chi \ne 0$ \cite{ss}. \end{theorem}
In the previous theorem $m({V_0})$ is  the sum of masses of vertices of $X_0$ (the $\mathbf{ cell\, mass\, of\, crystal}$).
\begin{figure}\label{2}
	
	%\centering
	
	{\includegraphics
		[width=13cm]{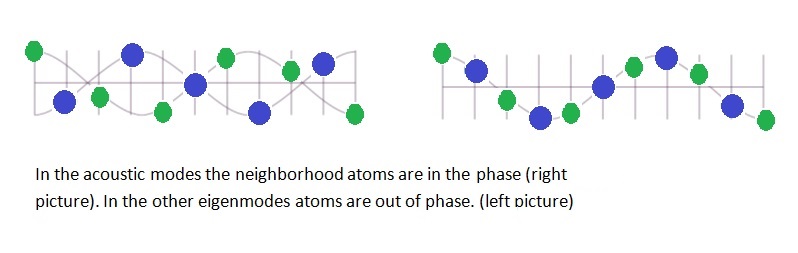}}
\end{figure}

\section{Laplacian on tori}
For simplicity denote $\rho(L)$ also by $L$. The character group of the lattice L is equal to the torus $\widehat{L}=\frac{\mathbb{R}^n}{L^*}$ where ${L^*}$ is the reciprocal lattice of L, i.e\\                   \begin{equation}L^*=\{x\,|\,x.y\in \mathbb{Z},\, \forall y\in L\}.\end{equation}\\ Equip each of these tori with the natural Euclidean metric. The length of  closed geodesics of  $\widehat{L}$ are the same as the eigenvalues of the Laplace-Beltrami Operator on the torus $\frac{\mathbb{R}^n}{L}$. Let us integrate $s_i(\chi)^2$ over simple closed geodesics including $1$. We call this the integrated acoustic velocity.
A natural question arising here is: \\\\ $\mathbf{What \,is\, the\, ralation\, between\, integrated \, acoustic\, velocity\, of\, the \, crystal\, lattice \,}$ $C$\,$\mathbf{ and \,the\, }$\\  $\mathbf{ \,Laplacian \,eigenvalues\, of\, the \,torus \, }$$\frac{\mathbb{R}^n}{L}$?
\subsection{Algebraic formalism}

Even though, there are many choices leading us to diverse nice problems about the acoustic phase velocity and the spectrum of the symmetric torus, in this subsection we only investigate the integration of $\sum_{i=1}^n s_i(\chi)^2$.  Let $ T$ denote the set of  all simple closed geodesics of $  \widehat{L}$ initiated from the identity and parametrized by the interval [0,1]. 
\begin{definition} The integrated acoustic phase velocity on $c\in T$ is defined by $\int_c\sum_{i=1}^n s_i(\chi)^2d\chi$. 
\end{definition}
\begin{definition} The $\mathbf{integrated\,acoustic \,spectrum}$ of a crystal lattice $C$ is defined by the set  \begin{equation} \{\int_c\sum_{i=1}^n s_i(\chi)^2d\chi|c \in T\}.\end{equation}
\end{definition} We may pay to the next question:\\
$\mathbf{When\, can\, we\, hear \,the\, spectrum\, of \, the\, torus}$ $\frac{\mathbb{R}^n}{L}$$\mathbf{\, from\, the  \,acoustic\, spectrum\, of \,the}$\\ $\mathbf{\, crystal\,}$$\mathbf{ lattice\, C?}\,\,(**)$
\begin{theorem}\label{nn} Consider m vectors $v_i=v(e_i), i=1,\cdots,m$ where $v(e_i)=\phi (t({e_i}))-\phi(e{(e_i)})$.  Assume $|v_i|=1$ and $trA(e_i)=\frac{
		3}{2\pi^2}m(V_0),\,\, i=1,\cdots,m$. Then the acoustic spectrum is equal with $Asp=\{\sum_i(\chi.v_i)^2 |\chi \in L\}$.
	\end {theorem}
	Proof: The therorem  is a result of the fact that  the summation $\sum_{i=1}^n s_i(\chi)^2$ is the trace of the matrix $A_\chi$(Theorem \ref{3}) and the fact that %Applying assumptions $\sum_{i=1}^n s_i(\chi)^2=\sum_i(\chi.v_i)^2$
	$ \int_c\sum_{i=1}^n s_i(\chi)^2=\sum_{i=1}\int_0^1s_i(t\chi)^2dt=\sum_{i=1}\int_0^1s_i(t\chi)^2dt=\int_0^1t^2s_i(\chi)^2dt=\frac{2\pi^2}{
		3m(V_0)}
	\sum_{e\in E_0}(\chi.v(e_i))^2trA(e_i)$.
	\\\\The problem $(**)$  can be written as follows,\\
	\hspace{2cm} Let $Asp=\{\sum_i(\chi.v_i)^2 |\chi \in L\}$,  can we determine the set $Lsp=\{|\chi| | \chi \in L\}?$ 
	\begin{theorem}\label{43} Under the assumptions of Theorem \ref{nn}, the $Asp$ determine the lengths of elements of $L$ and ${L}^*$ up to a constant $c$.  
	\end{theorem} Proof: According to the property \ref{se} of a standard realization, $\sum_i(\chi.v_i)^2=c|\chi|^2$. On the other hand from the Poisson formula we have  $ \sum_{y\in L^*}
	e^{-4\pi^2 {|y|}^2t}=\frac{Volume(L)}{{(4\pi t)}^\frac{n}{2}}\sum_{s\in L}
	e^{-\frac{{|s|}^2}{4t}}$ which provides the relation between lengths of elements of $L$ and $L^*$.\\
	The physical interpretation of Theorem \ref{43} is that when the average of quantities $(A(e)x.x)$ on the unit sphere (which is a divergence like quantity )  are the same for  neighboring atoms and the crystal laies in its harmonic position,  the integrating of velocities ($s_i(\chi)$) of independent acoustic phases over the closed geodesics of the angular momentum phase space determine the length of elements of the symmetric lattice of the crystal up to a constant. 
	\subsection{Generalized problem }
	In this  subsection we consider examples for a more general case.  We ignore that the set of vectors $v_i,\,\, i=1,...,n$ are obtained from a lattice harmonic embedding. Moreover we assume some extra assumptions about our lattice.\\
	Example 1: Let $v_1,v_2,v_3$ be an orthogonal basis of $\mathbb{R}^3$ and $v_4$ makes angle $120^\circ$ with  each vector $v_1,v_2,v_3$. Also let the position of $v_4$ is such that $\chi.v_4\in \mathbb{Q}^c$ for  $\chi \in L$. Then the set $Asp$ is equal to $\{|\chi|^2(1+cos^2\theta_{4\chi})|\chi \in L\}$ where $cos^2\theta_{4\chi}$ is irrational for all  $\chi \in L$ which denotes the angle between $\chi$ and $v_4$. We can obtain candidates for  $Lsp$'s by dividing each element of the set $Asp$ by the  numbers in the interval $(1,2)$ which generate  integer numbers and then computing the square root of them.\\\\
	Example 2:
	In this example let us conceive some extra knowledge about $L$ and the vectors $v_i,\,\,i=1,\cdots,n.$  Suppose  we know there are four basic atoms in each cell of a crystal and one of them is jointed with the others.%; This has been illustrated in the figure\ref{?????}.
	Therefore, we have three vectors and assume they have the same length 1.  Furthermore suppose that there is a number $k$ such that $kL$ is an integral lattice generated by two vectors of the same length. Concisely 
	let $L=\{k\chi +l\eta: k,l\in \mathbb{Z}\}$ and
	$|\eta|=|\chi|.$\\
	%3. $Asp=\{k^2|\chi|^2(\sum_{i=1}^3cos^2\theta_{i\chi}) +l^2|\eta|^2(\sum_{i=1}^3cos^2\theta_{i\eta})+2kl|\chi||\eta|(\sum_{i=1}^3cos\theta_{i\chi}cos\theta_{i\eta})|k,l\in \mathbb{Z}\}$\\
	By these conditions we have $$ Asp=|\chi|^2\{k^2(\sum_{i=1}^3cos^2\theta_{i\chi}) +l^2(\sum_{i=1}^3cos^2\theta_{i\eta})+2kl(\sum_{i=1}^3cos\theta_{i\chi}cos\theta_{i\eta})|k,l\in \mathbb{Z}\},$$ where $\theta_{i\chi}$ and $\theta_{i\eta}$
	are respectively  the angles of $v_i$ with $\chi$ and $eta$. Assume that  $\sum_{i=1}^3cos\theta_{i\chi}cos\theta_{i\eta}$ be a positive number, then the minimum  of the set $\frac{T}{|\chi|^2}$  is equal to  $$min\{(\sum_{i=1}^3cos^2\theta_{i\chi}) ,(\sum_{i=1}^3cos^2\theta_{i\eta}) \}.$$ Now , assume that we know all solutions of the next algebraic problem.\\
	Problem.Let $\mathbb{Z}*\mathbb{Z} =\{x^2|x\in \mathbb{Z}\}$ and $\mathbb{Z}. \mathbb{Z}=\{x.y |x,y\in \mathbb{Z}\}$.
	Suppose that $m\in \mathbb{Z}$, $\alpha,\beta,\gamma \in \mathbb{R}^+$, and assume that we know the set $M=\{m(\alpha \mathbb{Z}*\mathbb{Z}+\beta \mathbb{Z}*\mathbb{Z} +\gamma \mathbb{Z}. \mathbb{Z})\}$. Find all four tuple $(m,\alpha,\beta,\gamma)$ with the same $M$.\\\
	For each 3 tuples $(\alpha,\beta,\gamma)$ we must find the  set of  solutions for the set of equations  $$\sum_{i=1}^3cos^2\theta_{i\eta}=\beta,\sum_{i=1}^3cos\theta_{i\chi}cos\theta_{i\eta}=\gamma,\sum_{i=1}^3cos^2\theta_{i\chi}=\alpha.$$ A simple geometric discussion on the angles provide a description of the set $Lsp$. 

%%%%%%%%%%%%%%%%%%%%%%%%%%%%%%%%%%%%%%%%%%%%%%%%%%%%%%%%%%%%%%%%%%%%%%%%%%%%%%%%%%%%%%%%%%%%%%%%%%%%%%%%%%%%%%%%%%%%%%%%%%%%%%%%%%

\end{document}